\documentclass[preprint,aps,prd,amsmath,amssymb]{revtex4}
\usepackage{graphicx}
\usepackage{color}
\usepackage[latin1]{inputenc}
\newcommand{\be}{\begin{eqnarray}}
\newcommand{\ee}{\end{eqnarray}}

\newcommand{\bmp}{\noindent\begin{minipage}{16cm}}
\newcommand{\emp}{\end{minipage}\vskip 7mm} % 7mm untightened

\usepackage{graphicx}% Include figure files
\usepackage{dcolumn}% Align table columns on decimal point
\usepackage{bm}% bold math
\usepackage{amsmath}
\usepackage{amsfonts}
\usepackage{bbm}
\usepackage{subfigure}

\def\drawbox#1#2{\hrule height#2pt
        \hbox{\vrule width#2pt height#1pt \kern#1pt
              \vrule width#2pt}
              \hrule height#2pt}

\def\Asym#1#2{\vcenter{\vbox{\drawbox{#1}{#2}
              \kern-#2pt % line up boxes
              \drawbox{#1}{#2}}}}

\usepackage{amssymb}
\newcommand{\beq}{\begin{equation}}
\newcommand{\eeq}{\end{equation}}
\newcommand{\bea}{\begin{eqnarray}}
\newcommand{\eea}[1]{\label{#1}\end{eqnarray}}

\begin{document}
\thispagestyle{empty}

{\hfill \parbox{4cm}{
        SHEP-05-38 \\
}}

%%%%%%%%%%%%%%%%%%%%%%%%%%%%%%%%%%%%%%%%%%%%%%%%%%%%%%%%%%%%%%%%%%%%%%%%%%%
\title{Minimal Walking Technicolour, the Top Mass\\ and \\ Precision
Electroweak Measurements}
\author{Nick {\sc Evans}}
\email{evans@phys.soton.ac.uk}
\affiliation{School of Physics and Astronomy
Southampton University, Southampton, S017
1BJ, United Kingdom}
\author{Francesco {\sc Sannino}}
 \email{sannino@nbi.dk}
\affiliation{The Niels Bohr Institute, Blegdamsvej 17, DK-2100 Copenhagen \O, Denmark }

%%%%%%%%%%%%%%%%%%%%%%%%%%%%%%%%%%%%%%%%%%%%%%%%%%%%%%%%%%%%%%%%%%%%%%%%%%%%%%%%%%%%%%%%%%%%%%%%%%%%%%%%%%%%%%%%%%%%%%%%%%%%%%%%%%%%%%%%%%%%%%

\begin{abstract}
We consider a minimal technicolour theory with two techniflavours in
the adjoint representation of an SU(2) technicolour gauge group
which has been argued to feature walking dynamics. We show how to
naturally embed this theory in an extended technicolour model
capable of generating the top quark mass. We investigate the
precision constraints and conclude that such models, in the light
of the most recent precision data fit, are not ruled out.

\end{abstract}

%%%%%%
%%%%%%%%%%%%%%%%   TITLE    %%%%%%%%%%%%%%%%%%%%

\maketitle

%%%%%%%%%%%%%%%%%%%%%%%%%%%%%%%%%%%%%%%%%%%%%%%%%%%%%%%%%

\section{Introduction}

A dynamical mechanism for the breaking of electroweak symmetry, in
the spirit of the chiral symmetry breaking driven by QCD's
dynamics, has long been an appealing idea. Technicolour is the
prototype theory \cite{TC}. The light standard model fermion
masses must be produced by the introduction of interactions
linking those fermions to the techni-fermion sector. Here extended
technicolour (ETC), where the technicolour group is combined with
a gauged flavour symmetry of the standard model fermions, is the
archetype \cite{Lane:1989ej}. The massive ETC gauge bosons
generate light masses given by approximately  ${v^3/M^2_{ETC}}$.
Flavour physics has the potential to generate large flavour
changing neutral currents (FCNCs) for the second family although
this physics is at scales of at least 10s of TeV and hence hard to
probe. Many and varied models exist of these high ETC scales
including models with GIM mechanisms \cite{Georgi:1992at,
Randall:1992vt,King:1992qb,Burdman:1998vw} that survive flavour
bounds.

In recent years precision electroweak constraints (neatly
encapsulated by the measured values of the parameters S and T
\cite{Peskin:1990zt,Peskin:1991sw,Kennedy:1990ib,Altarelli:1990zd,Marciano:1990dp})
have exerted considerable pressure on such models, ruling out many
simple cases. These constraints have tended to favour the standard
model with a light higgs and zero or negative contributions to S
and T. Most technicolour models on the other hand give rise to a
heavy higgs and large positive S and T contributions. Indeed
generally mechanisms for generating negative S and T are thin on
the ground. The full set of data from the LEP, SLC and Tevatron
experiments is now available and a final set of limits on S and T
have been produced \cite{unknown:2005em}. Interestingly the
central value has shifted, as a result of more precise
measurements, and the data now shows some preference for positive
S and T contributions. We reproduce a plot of the latest fit
results in the S,T plane in Figure 1.
\begin{figure}[!htp]
\begin{center}
\includegraphics[scale=.5]{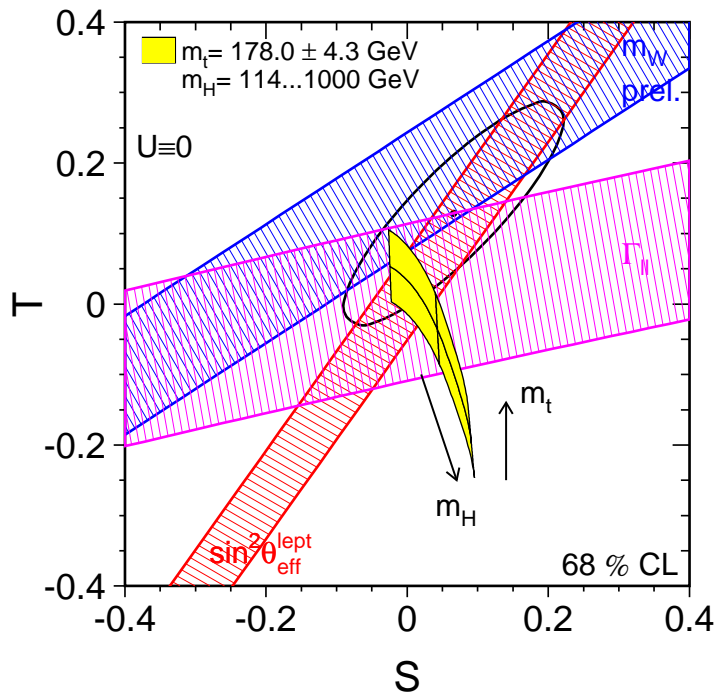}
\caption{Latest electroweak precision measurements taken from
\cite{unknown:2005em}. The ellipse is drawn for a reference higgs
mass of 150~GeV.} \label{qqgluonloop}
\end{center}
\end{figure}
 This data is more favourable to the generic idea of
there being physics beyond the standard model.

Much has been made of the difficulties technicolour models have
with S although the most minimal models \cite{Sundrum:1992xy}
(such as an SU(2) technicolour group and a single techni-fermion
doublet) have always been reconcilable with the data. The most
stringent hurdle to dynamical symmetry breaking models came
instead from the large top mass. Generating such a large mass
seems to require a very low ETC scale laying the mechanism open to
test. This ETC sector must incorporate large isospin breaking to
explain the top bottom mass splitting. The contributions to the T
parameter can be huge \cite{Chivukula:1995dc,Appelquist:1996kp}.
The simplest models of technicolour with QCD-like dynamics and a
naive ETC mechanism are undoubtable ruled out.

QCD has traditionally been used as the testing ground of
technicolour dynamics but it has always been recognized that
non-QCD-like dynamics might play a part. The most discussed
example is walking technicolour
\cite{Holdom:1981rm,Yamawaki:1985zg,
Appelquist:an,MY,Cohen:1988sq}. Such theories are assumed to lie
close to a strongly coupled infra-red fixed point so that the
gauge coupling is strong over a significant energy regime. This
behaviour enhances the size of the fermion condensate that is
assumed to form and break electroweak symmetry. This in turn
enhances the light fermion masses generated by ETC allowing the
ETC scale to be raised. Originally the walking mechanism was used
to address the FCNC problem requiring walking over many decades of
energy regime. This seems to ask a bit much of the near
conformality of the theory. Recently though the idea has been
resurrected to push the ETC scale associated with the top quark
higher to reduce T contributions \cite{Appelquist:2003hn}. Here
conformality must only be approximated over a few TeV energy
range. We will investigate this idea in a new setting below.

Another consequence of conformality is that the technifermion self
energy is enhanced at high energy. This reduces the contributions
to the S parameter
\cite{Appelquist:1998xf,Appelquist:1999dq,Sundrum:1991rf,Harada:2005ru}.

Finally it has been argued that walking theories are liable to
generate a lighter higgs than traditional technicolour
\cite{Dietrich:2005jn}. To trigger chiral symmetry breaking the
value of the coupling at the IR fixed point must lie close to a
critical coupling value. If the fixed point was above, chiral
symmetry breaking would be triggered before the fixed point was
reached, if it was too low, there would be no chiral symmetry
breaking. If we treat the number of techni-fermion flavours as a
continuous variable then by adjusting $N_f$ one should be able to
cross through this quantum phase transition. If one thinks in
terms of an effective sigma model description of the symmetry
breaking then, if the transition is continuous, it is reasonable
that at the critical value of $N_f$ the higgs mass falls to zero
\footnote{The mean field result for the higgs mass as function of
number of flavors found in \cite{Dietrich:2005jn} may acquire
corrections \cite{Appelquist:1991kn,Sannino:1999qe}. A possible
criticism would be that near the conformal phase transition other
states may become light (although no formal proof exists).
Although this may affect the argument supporting universal
behavior near the phase transition \cite{Chivukula:1996kg}, it
need not influence the result \cite{Dietrich:2005jn}. We also stress that the fact that the
chiral symmetry breaking scale vanishes as well does not imply that the chiral partner of the pions
does not become parametrically lighter and narrow near the phase transition. This is so since the
self coupling of the scalar is also a parameter which is expected to vanish
near a continuous phase transition.}. Thus if we
lie close to the critical value in order to achieve walking the
higgs mass might be expected to be lower than that in QCD.

Recently developments in the understanding of supersymmetric gauge
theories \cite{Intriligator:1995au} have revealed that strongly
coupled gauge theories can indeed have conformal IR behaviour for
some range of the number of quark flavours. Arguments at large $N$
have been used to connect these theories to non-supersymmetric
gauge theories with two index tensor matter \cite{Armoni:2003gp}
(such a connection does not extend to confinement properties
\cite{Sannino:2005sk}). In \cite{Sannino:2004qp} the implications
of these theories to technicolour were raised. Using some of the
results worked out in \cite{Sannino2003:xe} it was suggested, in
\cite{Hong:2004td}, that the composite higgs can be light in these
strongly coupled theories.

The electroweak data though forces one to propose small $N$
examples. The most minimal model proposed as having a conformal
fixed point is SU(2) with two adjoint fermions
\cite{Sannino:2004qp}. This is the model we will develop further
in this letter. We note that it is always possible to propose that
any technicolour model can be made walking by adding additional
appropriate matter multiplets. If these are electroweak singlets
then they will be invisible in precision data
\cite{Sundrum:1992xy,Christensen:2005cb}. It would be more elegant
though if the electroweak content were sufficient and we pursue
these examples since they are perhaps the most minimal matter
content theories with possible walking dynamics.

The idea of technicolour models with matter fields in higher
dimension representations of technicolour have been discussed
before \cite{Corrigan:1979xf,{Lane:1989ej}}. Such models are typically hard to combine with an extended
technicolour sector since the standard model fermions are in the
fundamental representation of their flavour symmetries \cite{Christensen:2005cb}. The
generation of the light fermion masses must be a crucial part of
an electroweak symmetry breaking model. Here we will show that in
fact the SU(2) model with adjoint matter can rather naturally be
recast so that it is accessible to a standard ETC scenario. The
crucial observation is that the model can also be written as an
SO(3) technicolour model with fundamental matter. We will
concentrate on the ETC mechanism for the generation of the top
mass since only that sector will be experimentally accessible by
the LHC. There is no block to using standard ETC technology to
generate the lighter fermion masses too though.

We will finally address the crucial issue of the impact of the
model on precision electroweak data. We estimate the contributions
to S and T. We argue that the contributions, including the effects
of the walking dynamics, are of order the allowed positive shifts
in S and T with a somewhat heavy higgs. These estimates are as
usual in technicolour only order of magnitude estimates but one
must conclude that the dynamics of this model is not clearly ruled
out. Thankfully the LHC will switch on soon and place such
speculation on its correct footing.

\section{The Minimal Walking Model}

The technicolour sector we will consider is an SU(2) technicolour
gauge group with two adjoint technifermions.

The two loop beta function (which is exact in 'tHooft gauge) of
the more general theory with $N$ colours and $N_f$ adjoint
fermions is given by

\beq
\beta= -\beta_0 \frac{g^3}{16\pi^2} - \beta_1 \frac{g^5}{(16\pi^2)^2} \ ,
 \eeq
 with
 \beq
 \beta_0  =\frac{N}{3}\left(11- 4\,N_f \right)\ , \qquad \beta_1 = \frac{N^2}{3}\left( {34} - 32 N_f \right)\ .
 \eeq
The theory is therefore asymptotically free if $N_f < 2.75$. Below
this number of flavours, at large $N$ there is a perturbative
Banks Zak fixed point. As $N_f$ is lowered further this fixed
point becomes non-perturbative.

To estimate the critical coupling for chiral symmetry breaking we
require that the anomalous dimension of the quark mass operator
must satisfy the relation $\gamma(2-\gamma)=1$ \cite{Cohen:1988sq}. This yields
\begin{equation}
\alpha_c  \simeq \frac{\pi}{3N}.
\end{equation}
The critical value of the number of flavors which gives this fixed
point value is
\begin{equation}
 N_f^c \simeq 2.075 \ .
\end{equation}
Since we are considering adjoint Dirac fermions the critical
number of flavors, at the level of the approximations used here,
is independent of the number of colors \footnote{We thank D.
Dietrich and K. Tuominen for pointing this feature to us.}. If we
trust this result we can work at large $N$ where the critical
coupling vanishes to determine the critical value of $N_f$ in a
weakly coupled theory where we can trust the perturbative
beta-function results. This makes the estimates of the conformal
window here more solid than in the usual case with fundamental
matter.

Finally we note that the critical coupling value for $N=2$ is
$\alpha_c \simeq 0.52$ \cite{Sannino:2004qp}. We expect that the
theory will enter a conformal regime unless the coupling rises
above the critical coupling triggering the formation of a fermion
condensate.

We conclude that a $N_f=2$ theory is so close to $N_f^c$ that it
stands a good chance of being a walking theory - in other words
the coupling might spend a large energy regime running just below
the critical coupling before eventually just crossing it. When
chiral symmetry breaking is triggered the fermions decouple from
the theory and the pure glue dynamics then rapidly reach very
strong coupling and generate a mass gap for the gluonic degrees of
freedom.

To check the pattern of chiral symmetry breaking and for our
analysis of ETC in the model it is helpful to recast the theory.
We note that SU(2) with adjoint matter fields is equivalent to an
SO(3) gauge theory with fundamental matter multiplets. Thus the
two adjoint fermions may be written as

\beq \left(  \begin{array}{c} U^{a} \\D^{a} \end{array}\right)_L ,
\hspace{0.5cm} U_R^a , \hspace{0.5cm} D_R^a  \qquad a=1,2,3 \eeq
with $a$ the color index of SO(3). The left fields are arranged in
three doublets of the SU(2)$_L$ weak interactions in the standard
fashion. The symmetry breaking condensate is expected to be the
usual QCD-like vev for $\langle \bar{U}U + \bar{D}D \rangle$ which
breaks the electroweak symmetry in the standard technicolour
pattern.

This model as described so far suffers from a topological anomaly
- the Witten anomaly \cite{Witten:fp}. An SU(2) gauge theory must
have an even number of fermion doublets to avoid the anomaly. Here
there are three extra electroweak doublets added to the standard
model and we are required to add a further doublet. We do not wish
to disturb the walking nature of the technicolour dynamics so the
doublet must be a technicolour singlet \cite{Dietrich:2005jn}. Our
additional matter then is essentially a copy of a standard model
fermion family with quarks (here transforming as a 3 of SO(3)) and
a lepton doublet. It looks natural to make the same hypercharge
assignments as we do across a normal fermion family (other
possibilities exist as discussed in \cite{Dietrich:2005jn} but we
will stick to this case here). Clearly this fourth family lepton
will need to be made massive and the generation of such masses and
the top quark mass will be the challenge for the next section.

\section{Extended Technicolour}

We have seen that adjoint multiplets of SU(2) can be written as
fundamental representations of SO(3). This trick will now allow us
to enact a standard ETC pattern from the literature - it is
particularly interesting that for this model of higher dimensional
representation techniquarks there is a simple ETC model. We have
seen that the extra doublets of the model fill out a fourth family
so it is tempting to try to enact a one family ETC type model. In
fact though the techni-quarks do not transform under colour (the
distinct SO(3) technicolour group replaces these interactions for
the fourth family) and these schemes will not work. Instead we
will follow the path proposed in \cite{Randall:1992vt} where we
gauge the full flavour symmetry of the fermions.

If we were simply interested in the fourth family then the
enlarged ETC symmetry is essentially a Pati-Salam type
unification. We stack the doublets

\beq \left[\left( \begin{array}{c} U^a \\ D^a \end{array}
\right)_L \ , \quad \left( \begin{array}{c} N \\ E \end{array}
\right)_L \right],\hspace{1.5cm} \left[U_R^a\ ,~N_R \right],
\hspace{1.5cm} \left[D_R^a \ ,~E_R \right]\eeq into 4 dimensional
multiplets of SU(4). One then invokes some symmetry breaking
mechanism at an ETC scale (we will not speculate on the mechanism
here)

\beq SU(4)_{ETC} \rightarrow SO(3)_{TC} \times U(1)_Y \eeq

The technicolour dynamics then proceeds to generate a techniquark
condensate $\langle \bar{U} U \rangle = \langle \bar{D} D \rangle
\neq 0$. The massive gauge bosons associated with the broken ETC
generators can then feed the symmetry breaking condensate down to
generate fourth family lepton masses

\beq m_N = m_E \simeq {\langle \bar{U}U \rangle \over
\Lambda_{ETC}^2} \eeq We will estimate this in more detail below.

One could now naturally proceed to include the third (second,
first) family by raising the ETC symmetry group to SU(8) (SU(12),
SU(16)) and a series of appropriate symmetry breakings. This would
generate masses for all the standard model fermions but no isospin
breaking mass contributions within fermion doublets. The simplest
route to generate such splitting is to make the ETC group chiral
so that different ETC couplings determine the isospin +1/2 and
-1/2 masses. Let us only enforce such a pattern for the top quark
and fourth family here since the higher ETC scales are far beyond
experimental probing.

We might for example have the SU(7) multiplets

\beq \left[\left( \begin{array}{c} U^a \\ D^a \end{array}
\right)_L\ , \quad \left( \begin{array}{c} N \\ E \end{array}
\right)_L\ , \quad \left( \begin{array}{c} t^c \\ b^c \end{array}
\right)_L \right], \hspace{1.5cm} \left[U_R^a\ ,~ N_R\ ,~t^c_R
\right] \eeq here $a$ will become the technicolour index and $c$
the QCD index. We also have a right handed SU(4) ETC group that
only acts on \beq \left[D_R^a\ , \quad E_R \right]\eeq

The right handed bottom quark is left out of the ETC dynamics and
only has proto-QCD SU(3) dynamics. The bottom quark will thus be
left massless. The symmetry breaking scheme at, for example, a
single ETC scale would then be

\beq SU(7) \times SU(4) \times SU(3) \rightarrow SO(3)_{TC} \times SU(3)_{QCD}  \eeq

The top quark now also acquires a mass from the broken gauge
generators naively equal to the fourth family lepton multiplet.
Traditional estimates of the electroweak vev and ETC generated
masses based on one loop diagrams are given by

\beq v^2 = {N \over 4 \pi^2} \Sigma(0)^2, \hspace{1cm} m = {N
\over 4 \pi^2} {\Sigma(0)^2 \over \Lambda_{ETC}^2} \Sigma(0) \eeq
here $\Sigma(0)$ is the techniquark self enery at zero momentum
which must lie around 1TeV. Such estimates suggest that a top
quark mass beyond a few 10s of GeV would be hard to achieve even
with an ETC scale of 1 TeV.

Walking dynamics has many features though that one would expect to
overcome the traditional small size of the top mass in ETC models.
Firstly the enhancement of the techniquark self energy at high
momentum enhances the ETC generated masses by a factor potentially
as large as  $\Lambda_{ETC}/ \Sigma(0)$. In
\cite{Appelquist:2003hn} it is argued that this effect alone may
be sufficient to push the ETC scale to 4 TeV and still maintain
the physical top mass.

In fact the dynamics of walking models is most likely yet more
complicated! The technicolour coupling is near conformal and
strong so the ETC dynamics will itself be quite strong at its
breaking scale which will tend to enhance light fermion masses
\cite{Evans:1994fb}. In our ETC model the top quark will also feel
the effects of the extra massive octet of axial gluon-like gauge
fields that may induce a degree of top condensation a l\`{a} top
colour models \cite{Miransky:1989nu,Hill:1994hp}. A precise
estimate of this strongly coupled system is beyond current
theoretical understanding. Gap equation analyses in NJL type
models do support the enhancement. Note that because there is a
standard ETC mechanism present that generates a basic contribution
to the top mass the NJL type interactions do not have to be tuned
to criticality to provide some enhancement of the top mass. In any
case such a tuning is already implicit in the walking mechanism.

We conclude that a 4-8 TeV ETC scale for generating the top mass
does seem possible and is certainly an interesting possibility. In
this model the fourth family lepton would then have a mass of the
same order and well in excess of the current search limit $M_Z/2$.

\section{Precision Data}

We next come to estimating the contributions of the technifermion sector
of the minimal walking model to the S and T electroweak precision data parameters.

There are four extra massive electroweak doublets in this model - the three SO(3)
colours of techni-quarks and the extra lepton doublet. Assuming these doublets
are degenerate then one would perturbatively expect that there is an S parameter
contribution

\beq S \simeq 4 \times { 1 \over 6 \pi} \simeq 0.2 \eeq

If the technicolour dynamics were QCD-like then the contribution
from the techniquarks would be expected to double due to the
exchange of techni-glue \cite{Peskin:1990zt}. However, with
walking dynamics the techniquark self energies are expected to
fall off less quickly and the naive perturbative computation with
a hard mass is expected to be a conservative estimate. Indeed, one
can show that the S value in a walking theory is reduced with
respect to the naive perturbative estimate near the conformal
window \cite{Appelquist:1998xf,Sundrum:1991rf}.

The ETC gauge bosons in the model break custodial isospin to
generate the top bottom mass splitting and one must worry that
this will feed into the techniquark sector generating large T
contributions \cite{Chivukula:1995dc,
Appelquist:1996kp}. There are
a number of such contributions. Firstly one can consider the
exchange of an ETC gauge boson (associated with a diagonal
generator) across a techni-quark loop - in our model there are
such bosons that couple to $Q_L$ and $U_R$ but not $D_R$. One can
naively estimate these contributions from perturbative diagrams
but the result is essentially just a dimensional estimate

\beq T \sim {v^2 \over \alpha \Lambda^2_{ETC}} \sim 0.3 \eeq Here
$\alpha=e^2/4\pi$, we have used an ETC scale of 5~TeV and there
will also be order one multiplicative factors.

A second contribution comes from the exchange of ETC gauge bosons
enhancing the techni-quark masses. Again since there is isospin
breaking we might expect techni-up techni-down mass splitting. If
we use the perturbative estimate for the resulting contribution to
T we find \beq T \sim {N_{TC} \over 12 \pi^2 \alpha }{\Delta
\Sigma^2 \over v^2} \sim 0.6 N_{TC} {\Delta \Sigma^2 \over m_t^2}
\eeq Whether the techniquark mass splitting ($\Delta \Sigma^2$) is
actually as large as the top bottom mass splitting depends on the
precise origin of the top mass. For example if the top colour
interactions play a large role, the splitting in the techni-sector
may be smaller.

Finally there is the fourth family lepton multiplet. In the ETC model above if we
keep the SU(7) and SU(4) ETC groups couplings equal then this sector will
only see isospin splitting as a result of the top bottom splitting. The
lepton doublet might reasonably be expected to have the same mass splitting as
the techni-quark doublet shifting $N_{TC} \rightarrow N_{TC}+1$ in the above
estimate.

These estimates are rather naive and the most we can claim is that
the T contribution will be of order 0.5 multiplied by a number of
order one. If we now refer back to Figure 1 we see that if the
walking dynamics produces a higgs with mass between 150~GeV to
1~TeV and we allow an S contribution of 0.2 and a T contribution
of 0.5 we are consistent with the data (note that only the shift
in the latest precision data allows such a conclusion). We
conclude that the model is not ruled out.

\section{Summary}

It has become clear from the study of supersymmetric and large N
gauge dynamics that theories with strongly coupled IR fixed points
do exist. A minimal version of such a theory might provide a
viable walking technicolour model. Here we considered a minimal
model SU(2) with two adjoint fermions that has been suggested as a
walking theory. Our main goal was to show that it can be naturally
extended to an ETC model. Only when the top mass generation
mechanism is explicit can one even begin to test whether the model
has too much isospin violation and hence too large a T parameter -
the biggest constraint on technicolour models. Our naive estimates
suggest that the T parameter contributions are of the order of
those now allowed by the precision data. We would caution that
such models can not yet be ruled out and that the latest precision
measurements have alleviated some of the stress these models are
under.

\end{document}